# Mechanics and Resonance of the Cyanobacterial Circadian Oscillator

Ioannis G. Karafyllidis
*Democritus University of Thrace,
67100 Xanthi, Greece.
e-mail: ykar@ee.duth.gr*

*Abstract:* Recent experiments elucidated the structure and function of the cyanobacterial circadian oscillator, which is driven by sunlight intensity variation and therefore by Earth's rotation. It is known that cyanobacteria appeared about 3.5 billion years ago and that Earth's rotational speed is continuously decreasing because of tidal friction. What is the effect of the continuous slowdown of Earth's rotation on the operation of the cyanobacterial oscillator? To answer this question we derived the oscillator's equation of motion directly from experimental data, coupled it with Earth's rotation and computed its natural periods and its resonance curve. The results show that there are two resonance peaks of the "cyanobacterial oscillator-rotating Earth" system, indicating that cyanobacteria used more efficiently the solar energy during the geological period in which the day length varied from about 11 to 15 hours and make more efficient use of solar energy at the geological period which started with a day length of 21 hours and will end at a day length of 28 hours.

## 1. Introduction

Earth's rotation generates a periodically fluctuating environment in which organisms have to survive and evolve. Several species developed molecular time-keeping mechanisms, called circadian oscillators or clocks, to anticipate this environmental fluctuation and coordinate metabolism and gene expression (Imaizumi *et al.*, 2007). The metabolism of autophototrophic cyanobacteria is greatly dependent on the alteration of day and night and they are the simplest organisms that developed an internal circadian oscillator (Kondo and Ishiura, 1999). The cyanobacterial circadian oscillator comprises three proteins: KaiA, KaiB and KaiC (Nakajima et al., 2005). When these proteins are mixed with ATP in a test tube they generate robust and sustained oscillations, which are expressed as the oscillation of the total amount of phosphorylated KaiC (Nishiwaki *et al.* 2007; Ito *et al.* 2007). These circadian oscillations stem from the ordered phosphorylation of KaiC at the serine 431 (S) and threonine 432 (T) sites (Rust e*t al.*, 2007). The cyanobacterial circadian oscillator is driven by light and dark pulses during its operation result in phase shifts (Rust *et al.*, 2011).

The Moon formed, about 4.5 billion years ago and since then, Earth's rotational speed decreases with time because of tidal friction (Canup and Asphaug, 2001; Sonett *et al.*, 1996). Several organisms such as corals, bivalves and cephalopods record in their skeletons the astronomical cycles of the Sun-Earth-Moon system and from these records the number of days per month and the number of months in a year can be obtained as functions of time (Sonett *et al.*, 1996; Sisterna and Vucetich, 1994) . Fossil record data and astronomic calculations show that the length of the terrestrial day increases about 20 ms every one thousand years (Sisterna and Vucetich, 1994).

Cyanobacteria appeared about 3.5 billion years ago and their circadian oscillator has been driven by day-night alterations the period of which has varied from about 6 to 24 hours. Every oscillator whether it is mechanical, electrical, molecular or quantum, has a natural frequency, which corresponds to a natural period. If the oscillator is not simple, it may have more than one natural periods. In the experiments of Rust et al. (Rust *et al.*, 2007) the structure and function of the cyanobacterial oscillator was studied *in vitro* with ATP in excess at all times, which corresponds to oscillations under constant illumination conditions throughout the duration of the experiments (Rust *et al.*, 2011). It is therefore possible that the information about the natural period(s) of the cyanobacterial oscillator is hidden in these experimental data. Furthermore, since this oscillator is driven by a periodic cause (Earth's rotation), the period of which varies with time, it would be interesting to compute the resonance curve of the "cyanobacterial oscillator-rotating Earth" system and associate it with geological time and events.

**2. Resonances of the cyanobacterial oscillator-rotating Earth system**

With their experiments, Rust *et al.* revealed the details of the molecular cyanobacterial oscillator (Rust *et al.*, 2007). The basic mechanism is the ordered phosphorylation and dephosphorylation of KaiC at both T and S sites, which is enhanced by KaiA. First, the amount of KaiC phosphorylated at the T site (T-KaiC) increases and shortly after that the amount of KaiC phosphorylated at both T and S sites (ST-KaiC) increases. When the concentration of ST-KaiC becomes high, KaiC starts to autodephosphorylate the T site and the amount of KaiC phosphorylated only at the S site (S-KaiC) increases. With the help of KaiB, S-KaiC sequesters KaiA and without it, KaiC dephosphorylation predominates. KaiC dephosphorylation causes S-KaiC concentration to decrease, KaiA is released, KaiC begins to phosphorylate again, the amount of T-KaiC increases and so on.

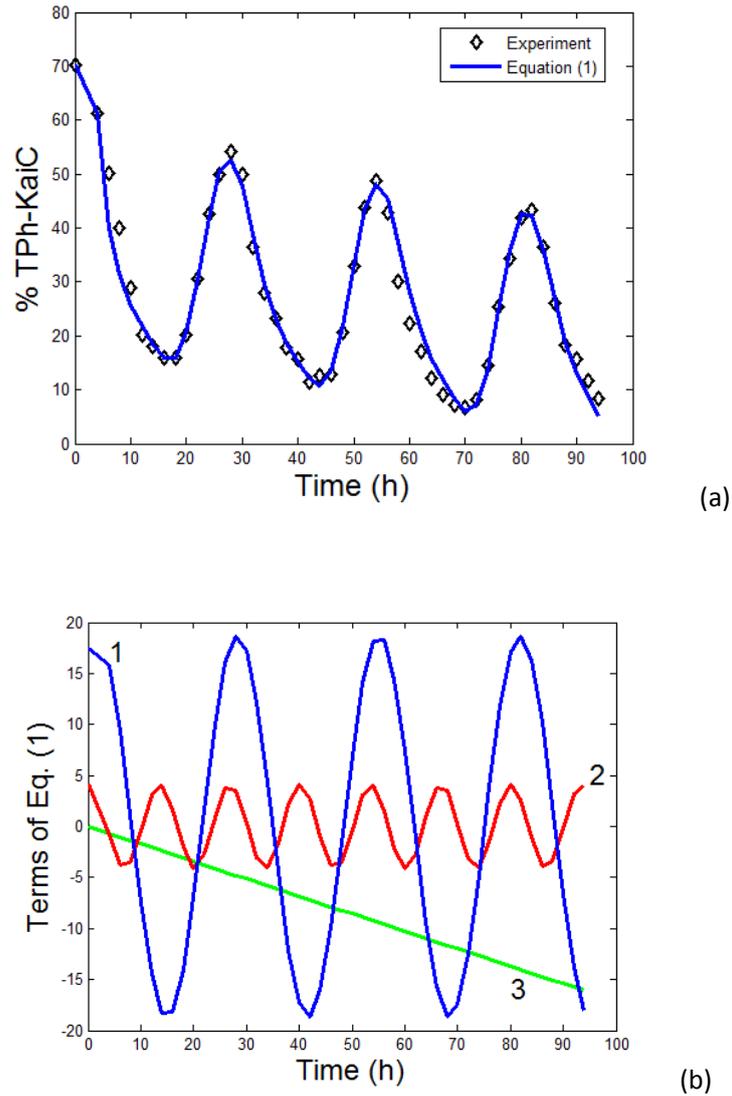

**Fig. 1.** (Color online) Time variation of TPh-KaiC. (a) Measurements of Rust *et al.* (◊ symbols) and time variation given by equation (1) (sold line). (b) Terms of equation (1). Term [*-0.171t*] with green, [*18.7 cos(0.235t+2.77)*] with blue and [*4.12 sin(0.472t+1.45)*] with red line.

This ordered phosphorylation combined with the negative feed-back loop created by S-KaiC, KaiB and KaiA results in the circadian oscillation of the total amount of phosphorylated KaiC. The measurements of Rust *et al*. of the total phosphorylated KaiC (TPh-KaiC) variation with time are shown in Fig. 1(a).

Schmidt and Lipson developed a method to obtain the natural laws that describe dynamical systems directly from experimental data and without any previous knowledge about the physics or the structure of the system (Schmidt and Lipson, 2009). Based on their method they developed a software tool and used it to find the equations of motion for simple and complex oscillating systems. We used the method of Schmidt and Lipson to obtain the equation that gives the variation of TPh-KaiC as a function of time. This equation is:

$$C(t) = 35.2 + 4.12 \sin(0.472t + 1.45) - 18.7 \cos(0.235t + 2.77) - 0.171t \qquad (1)$$

where C(t) represents the time variation of TPh-KaiC. Fig. 1(a) also shows the variation of TPh-KaiC given by this equation, which reproduces the experimental results with a mean absolute error equal to 0.12. Eq. (1) comprises three terms and a constant. These terms are shown in Fig. 1(b). As explained later on, the term (- 0.171 t) describes the slow decay of the oscillator proteins. The term, 4.12 sin(0.472 t + 1.45) describes a simple oscillation with period equal to 13.312 hours and stems probably from the phosphorylation and dephosphorylation of KaiC at the T site. The term, 18.7 cos(0.235 t + 2.77) is also a simple oscillation with period equal to 26.737 hours and stems probably from the phosphorylation and dephosphorylation of KaiC at the S site.

Since in the experiment of Rust *et al.* the amounts of T-KaiC and ST-KaiC rise and peak almost simultaneously, the phosphorylation at T and S states increases with about the same rate and the larger period of this term probably arises from the slow dephosphorylation of S-KaiC, because of the negative feed-back loop formed by S-KaiC, KaiB and KaiA. The experiment (Rust *et al.*, 2007) was performed with KaiB in excess and it is therefore unknown what effect the limited availability of KaiB would have on the period of the third term. Eq. (1) suggests that since KaiB is the key component in the feed-back loop, it is not entirely improbable that cyanobacteria are able to adjust their circadian rhythms, to some extent, by varying the concentration of KaiB.

For oscillating systems, the equation of motion is the equation that gives the second derivative with respect to time of the oscillating variable. The oscillating variable for the cyanobacterial circadian oscillator is the total amount of TPh-KaiC.

The second derivative of eq. (1) corresponds to the acceleration with which the oscillator changes state and is:

$$\frac{d^2 C(t)}{dt^2} = -a_1 \omega_1^2 \cos(\omega_1 t + \varphi_1) - a_2 \omega_2^2 \sin(\omega_2 t + \varphi_2) \qquad (2)$$

where $\omega_1=0.235$, $\omega_2=0.472$, $a_1=18.7$, $a_2=4.12$, $\varphi_2=1.45$ and $\varphi_1=\pi+2.77=5.912$. Setting:

$$x(t) = \cos(\omega_1 t + \varphi_1) \quad \text{and} \quad y(t) = \sin(\omega_2 t + \varphi_2) \qquad (3)$$

eq. (2) becomes:

$$\frac{d^2(x+y)}{dt^2} = \frac{d^2x}{dt^2} + \frac{d^2y}{dt^2} = -a_1\omega_1^2 x - a_2\omega_2^2 y \tag{4}$$

where for simplicity the parentheses that indicate the time dependence of x and y were omitted. Eq. (4) is the equation of motion of the cyanobacterial circadian oscillator.

The fact that a term originating from (-0.171 t) does not appear in the equation of motion, means that it does not describe any part of the oscillator. In other words it is not a damping factor but it rather stems from the slow decay of the oscillator proteins. This term and the constant 35.2 are the integration constants, which arise when eq. (4) is integrated twice to give the time variation of the oscillating variable.

The circadian oscillation is not damped because it is sustained by continuous consumption of ATP. The linearity of the equation of motion indicates that the cyanobacterial oscillator comprises two simple independent oscillators, one with period equal to 13.312 hours and one with period equal to 26.737 hours. It is reasonable to assume that since cyanobacteria experienced a day length variation that ranged from about 6 hours to 24 hours, they developed first an oscillator with a smaller period and then as the day length increased they evolved this oscillator to cope with longer days. The equation of motion indicates that the present molecular oscillator mechanism incorporates the older one.

In another experiment, Rust *et al.* showed that the amount of phosphorylated KaiC depends on the ADP/(ATP+ADP) ratio (Rust *et al.*, 2011). In the case of an ADP/(ATP+ADP) ratio equal to 75%, the final amount of phosphorylated KaiC saturates at about 50%, in the case of an ADP/(ATP+ADP) ratio equal to 50% it saturates at about

62% and when only ATP is present phosphorylated KaiC saturates at about 75%. This experiment showed that the amplitude of the circadian oscillation depends on the ATP concentration which cyanobacteria synthesize using solar radiation energy. Therefore the energy that drives the oscillator comes from sunlight and solar light intensity plays the same role that mechanical forces play in mechanical oscillators.

Solar energy is not available during the whole length of a terrestrial day, because of Earth's rotation. It is not available at night and its intensity is small in the morning hours, peaks at noon and becomes small again in the evening hours. Therefore the cyanobacterial circadian oscillator is driven by an energy source that is periodic but not sinusoidal. Furthermore, the period of that source is not constant but varies with geological time.

ATP provides the energy that drives the cyanobacterial circadian oscillator. Cyanobacteria synthesize ATP using solar radiation. Solar radiation intensity is not constant but varies as the Earth rotates around its axis. Therefore the period of the driving cause of the oscillator is equal to the period of Earth's rotation. Solar engineers developed very effective models to compute the daily variation of solar radiation intensity. One of the simpler and most useful models describes the variation of solar energy as a rectified sinusoidal function (Goswami *et al.*, 2000). The Fourier series expansion of this function is well known. The amplitude of this periodic function is larger at places near the equator and smaller at places near the poles, but the period remains constant at all places. After that, the solar light radiation intensity is given by:

$$L(t) = \Phi \left( 0.50 \sin(\omega t) - 0.21 \cos(2\omega t) - 0.04 \cos(4\omega t) \right) \qquad (5)$$

where Φ is power per unit area and equals to 930 W/m². Φ is the solar radiation intensity on Earth's surface when the Sun is found at the zenith. It is also known as air mass 1 (AM1) solar constant. Cyanobacteria use a part of this solar energy to synthesize ATP and we assume that the rate of ATP synthesis is proportional to solar radiation intensity and call *s* the proportionality constant. The oscillator consumes ATP and the consumption rate is proportional to the rate at which the concentration of TPh-KaiC increases (Terauchi *et al.*, 2007).

The amount of ATP that contributes to the ADP/(ATP+ADP) ratio which drives the oscillation is the difference between the ATP synthesis and consumption rates (Rust *et al.*, 2011) and is given by:

$$A(t) = s\, L(t) - g\, \frac{d(x+y)}{dt} \tag{6}$$

where A(t) is the time variation of the aforementioned ATP amount, (x+y) is TPh-KaiC and g is the proportionality constant that relates ATP consumption to TPh-KaiC variation and is estimated to be 0.05 mM/24h (Terauchi *et al.*, 2007).

After that, the equation of motion of the cyanobacterial oscillator driven by Earth's rotation is:

$$\frac{d^2x}{dt^2} + \frac{d^2y}{dt^2} + a_1 \omega_1^2\, x + a_2 \omega_2^2\, y = s\, L(t) - g\left(\frac{dx}{dt} + \frac{dy}{dt}\right) \tag{7}$$

in which the left hand side describes the time motion of the oscillating variable and the right hand side the cause that drives the oscillation.

Eq. (7) has the same form with the equation that describes a forced oscillation and taking into account the linearity of the equation, its resonance curve is given by:

$$\rho^2 = \frac{1}{(1/0.5\,s)^2 \left[\left(\omega^2 - a_1\omega_1^2\right)^2 + g^2\omega^2\right]} + \frac{1}{(1/0.5\,s)^2 \left[\left(\omega^2 - a_2\omega_2^2\right)^2 + g^2\omega^2\right]} +$$

$$\frac{1}{(1/0.21\,s)^2 \left[\left((2\omega)^2 - a_1\omega_1^2\right)^2 + g^2\omega^2\right]} + \frac{1}{(1/0.21\,s)^2 \left[\left((2\omega)^2 - a_2\omega_2^2\right)^2 + g^2\omega^2\right]} +$$

$$\frac{1}{(1/0.04\,s)^2 \left[\left((4\omega)^2 - a_1\omega_1^2\right)^2 + g^2\omega^2\right]} + \frac{1}{(1/0.04\,s)^2 \left[\left((4\omega)^2 - a_2\omega_2^2\right)^2 + g^2\omega^2\right]}$$

(8)

After that, we computed the resonance curve of the cyanobacterial circadian oscillator, which is shown in Fig. 2.

The quantity $\rho^2$ is proportional to the square of the amplitude of the oscillation and represents the energy that the driving cause transfers to the oscillator, which becomes larger as the period of the cause reaches the natural periods of the oscillating system and peaks when it becomes equal to these periods.

Since the cyanobacterial circadian oscillator regulates metabolism and gene expression, increasing oscillation amplitude means that cyanobacteria make more efficient use of solar energy. It is therefore reasonable to assume that, according to Fig. 2, cyanobacteria used more efficiently the solar energy during the geological period in which the day length varied from about 11 to 15 hours and make more efficient use of solar energy at the geological period which started with a day length of 21 hours and will end at a day length of 28 hours.

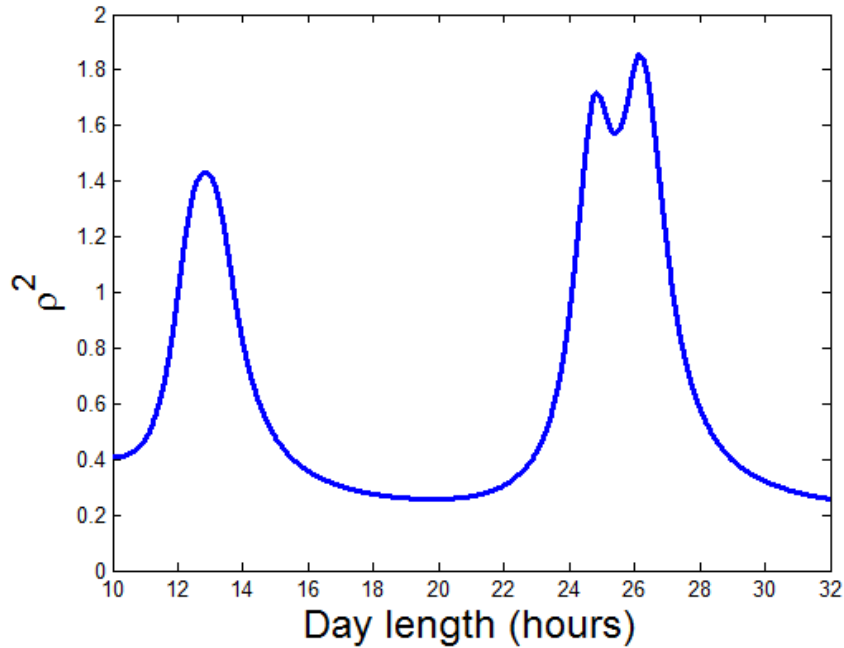

**Fig. 2.** (Color online). Resonance curve of the cyanobacterial circadian oscillator. In the x-axis, the angular frequency of Earth's rotation, $\omega$, has been replaced with the corresponding period ($\tau=2\pi/\omega$), which is the day length in hours.

Oxygen production is a result of cyanobacteria metabolism and the amount of oxygen produced by the cyanobacteria species during geological time is an indication of the efficient use of solar energy. Therefore, one expects higher oxygen levels in Earth's atmosphere at geological periods near the resonance peaks. The great oxidation event and the Cambrian explosion are two events that have a possible relation to increased oxygen production. The great oxidation event happened about 2.3 to 2.4 billion years ago when the day length was about 11 hours and Cambrian explosion happened about 530 to 550 million years ago when the day length was about 21 hours. Both events happened at about the onset of the two resonance peaks, when cyanobacteria were making more and more efficient use of solar energy. Although these two events are probably due to a combination of many factors, the resonances of the "cyanobacterial oscillator-rotating Earth" system may have had some contribution.

## 3. Conclusion

Recent experiments elucidated the structure and function of the cyanobacterial circadian oscillator. We now know enough about this oscillator to take the next step and couple it with its driving cause, Earth's rotation. There is a good reason for this. Cyanobacteria appeared about 3.5 billion years ago and since then Earth's rotational speed is continuously decreasing because of tidal friction. We tried to investigate the effect of the continuous slowdown of Earth's rotation on the operation of the cyanobacterial oscillator and on oxygen production. We constructed the equation of motion of the cyanobacterial oscillator directly from experimental data using a well-known method, coupled it with Earth's rotation and computed the resonance curve of the "cyanobacterial oscillator-rotating Earth" system. From this resonance curve one can see that the great oxidation event and the Cambrian explosion occurred at the onset of the two resonance peaks. It is therefore possible that the resonances of "cyanobacterial oscillator-rotating Earth" system may be related to those events.